\documentclass[aps,prl,10pt,twocolumn,floatfix,showpacs,superscriptaddress,longbibliography]{revtex4-1}

\usepackage{graphicx,amsmath,amsfonts,bm}
\usepackage[colorlinks=true, citecolor=blue, linkcolor=blue, urlcolor=blue]{hyperref}
\usepackage[all]{hypcap}
\usepackage{bbold}
\usepackage{ulem}

\begin{document}

\title{From fractional solitons to Majorana fermions in a paradigmatic model of topological superconductivity}
\author{N. Traverso Ziani}
\email{traversoziani@fisica.unige.it}
\affiliation{Dipartimento di Fisica, Universit\`a di Genova, 16146 Genova, Italy}
\affiliation{CNR spin, 16146 Genova, Italy}
\author{C. Fleckenstein}
\affiliation{Institute of Theoretical Physics and Astrophysics, University of W\"urzburg, 97074 W\"urzburg, Germany}
\author{L. Vigliotti}
\affiliation{Dipartimento di Fisica, Universit\`a di Genova, 16146 Genova, Italy}
\author{B. Trauzettel}
\affiliation{Institute of Theoretical Physics and Astrophysics, University of W\"urzburg, 97074 W\"urzburg, Germany}
\affiliation{W\"urzburg-Dresden Cluster of Excellence ct.qmat, Germany}
\author{M. Sassetti}
\affiliation{Dipartimento di Fisica, Universit\`a di Genova, 16146 Genova, Italy}
\affiliation{CNR spin, 16146 Genova, Italy}
\begin{abstract}
Majorana bound states are interesting candidates for applications in topological quantum computation. Low energy models allowing to grasp their properties are hence conceptually important. The usual scenario in these models is that two relevant gapped phases, separated by a gapless point, exist. In one of the phases, topological boundary states are absent, while the other one supports Majorana bound states. We show that a customary model violates this paradigm. The phase that should not host Majorana fermions supports a fractional soliton exponentially localized at only one end. By varying the parameters of the model, we describe analytically the transition between the fractional soliton and two Majorana fermions. Moreover, we provide a possible physical implementation of the model. We further characterize the symmetry of the superconducting pairing, showing that the odd-frequency component is intimately related to the spatial profile of the Majorana wavefunctions. 
\end{abstract}
\pacs{74.45.+c, 71.10.Pm, 74.20.Rp, 74.78.Na}

\maketitle

The search for platforms enabling the implementation of operations based on Majorana bound states is a fascinating area in condensed matter physics\cite{maj1,maj2,maj3,maj4,maj5}. Such devices represent a substantial step forward for topological quantum computation\cite{mqcanyons1,mqcanyons2}. As of now, the most promising candidates as hosts for Majorana bound states appear to be spin-orbit coupled quantum wires\cite{maj3,maj4,maj5}, planar Josephson junctions\cite{planar1,planar2}, topological insulators\cite{missings,aw}, and ferromagnetic chains on superconductors\cite{fmm}. The experimental tools commonly used to substantiate the formation of Majorana bound states in those systems are transport measurements and tunneling spectroscopy. A downside of such detection methods is that it is not easy to discriminate between topological Majorana bound states and trivial Andreev bound states\cite{tbs1,tbs2}, disorder\cite{dis1,dis2}, or distracting effects in Josephson junctions\cite{jj1,jj2,jj3,jj4}. More refined experimental schemes, involving for instance the study of non-local conductance\cite{nonloc} and current noise\cite{noise}, have hence been suggested to better characterize the presence of Majorana fermions. As the complexity of the properties to be inspected increases, the adoption of low energy models becomes more important to capture the essential physics.

A common trait of most low energy models for Majorana bound states is that they resemble the Jackiw-Rebbi model\cite{jr} in particle-hole space\cite{maj1}. The Majorana bound states are then located at mass kinks of the model. A competing topological bound state is naturally present in such models. When a spin (or chirality) index is also present, fractional solitons\cite{fr1,fr2,fr3,fr4} can emerge. These topological boundary states carry stable fractional charge\cite{stable} and have been predicted to appear in heterostructures based on topological insulators and ferromagnetic insulators\cite{fr1,fr2} or quantum point contacts\cite{fr4}. In lattice models, they arise in Su-Schrieffer-Heeger (SSH) like systems\cite{ssh}. They are fundamentally interesting and have been proven to lead to phenomena that can be potentially useful in spintronics\cite{fr1}. However, they have never been detected in a solid state setup.

Previously, the competition between phases hosting Tamm-Shockley\cite{pm1} states and Majorana fermions have been predicted in models based on spin-orbit coupled quantum wires\cite{fr3}. In that case, the appearence of zero modes when a single termination is imposed, has been analyzed.

In this work, we describe a simple superconducting system undergoing a transition between a state characterized by the presence of a single fractional soliton to a state characterized by two Majorana bound states. Our model generalizes the basic idea of a competition of these bound states invented in Ref.\cite{fr3} . The model is fully solvable with periodic and open boundary conditions at two ends. When periodic boundary conditions are imposed, a quantum phase transition between gapped phases is present. Unexpectedly, when open boundaries are considered, we show that a strictly zero energy solution is \textit{always} present. In one phase, the solution is localized at one end of the structure, in the other phase, it is located at both ends. The first case, being adiabatically connected to $\Delta=0$, corresponds to a single fractional soliton, the second to two Majorana zero modes. We interpret the result in terms of a heterostructure based on the helical edge states of a two-dimensional topological insulator proximitized by an s-wave superconductor\cite{str1,str2,str3,str4,str5,str6,str7,str8,str9,str10,str11,str12,str13,str14,str15,str16,str17,str18,str19}. If we inspect the Majorana phase in more detail, we are able to show a deep connection between Majorana wavefunction, tunneling density of states, and odd-frequency component of the anomalous Green function.

\begin{figure}
	\includegraphics[scale=1]{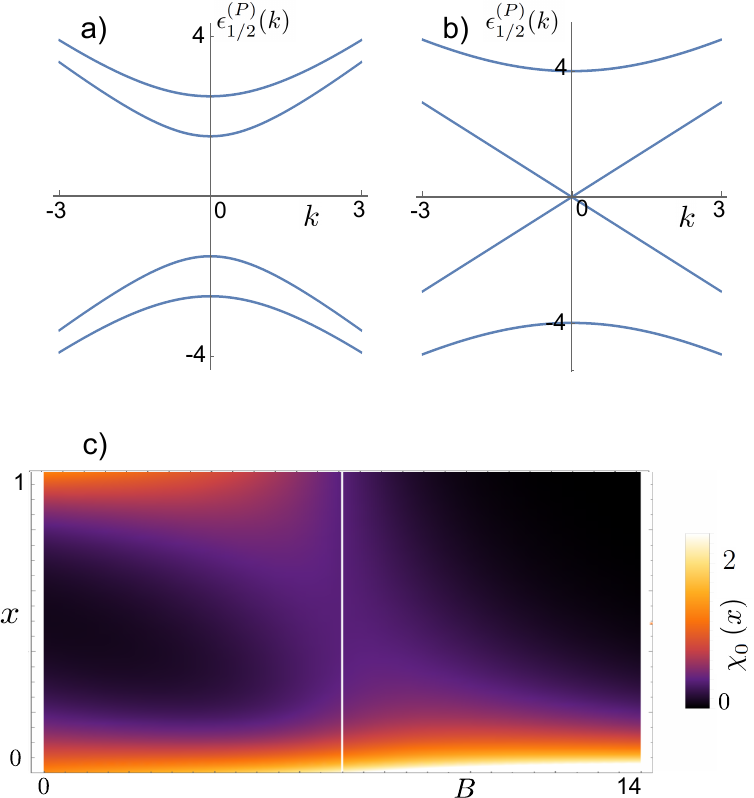}
	\caption{a) The dispersion $\epsilon^{(P)}_{1/2}(k)$ in units of $v_F/L$, as a function of $k$, in units $1/L$, for $\Delta=2v_F/L$ and $B=0.5v_F/L$.  b) The dispersion $\epsilon^{(P)}_{1/2}(k)$ in units of $v_F/L$ as a function of $k$ in units $1/L$, for $\Delta=2v_F/L$ and $B=2v_F/L$. c) Plot of $\chi_0(x)$, in units $L^{-1/2}$, as a function of $B$ in units $v_F/L$ and of $x$, in units $L$, for $\Delta=7v_F/L$. The central white line corresponds to the gapless point.}
	\label{Fig:map}
\end{figure}
The Bogoliubov-de Gennes (BdG) Hamiltonian of the model, on the segment of length $L$, that we study is ($\hbar=1$)
\begin{equation}
H=\frac{1}{2}\int_0^L \Psi^\dag(x)\mathcal{H}(x)\Psi(x)dx,
\end{equation}
where $\Psi^\dag(x)=(\psi_R^\dag(x),\psi_L^\dag(x),\psi_L(x),-\psi_R(x))$, with $\psi_{R/L}(x)$ Fermi operators, and the Hamiltonian density
\begin{equation}
\mathcal{H}(x)=-iv_F\partial_x \tau_z\otimes\sigma_z-B\tau_0\otimes\sigma_y-{i\Delta}\tau_x\otimes\sigma_0.
\label{eq:H}
\end{equation}
In Eq.(\ref{eq:H}), $v_F$ is the Fermi velocity, $B$ and $\Delta$ are real and positive competing masses of the model. Moreover, $\tau_i/\sigma_i$ are Pauli matrices acting respectively on particle-hole and $R/L$ space. Importantly, this Hamiltonian emerges, for instance, as a linearized model of a spinless topological superconductor (see the Supplememtary Material (SM)) at large chemical potential. Imposing periodic boundary conditions $\Psi^\dag(x)=\Psi^\dag(x+L)$, it is easy to obtain the spectrum of the Hamiltonian, given by the four excitation energy bands $\epsilon^{(P)}_{1/2}(k)=\pm\sqrt{v_F^2k^2+\Delta^2+B^2\pm 2\Delta B}$, where, $k=2\pi n/L$, with $n$ integer, represents the momentum eigenvalues. The dispersion relation is always gapped except  for $B=\Delta$. Moreover, it is even under the exchange of $B$ and $\Delta$ (see Fig.1a), b)). Differently from the case of spin-orbit coupled quantum wires\cite{soc,meng}, the model only has two Fermi points in the absence of masses. Hence, the naive expectation would be that, in case of open boundary conditions, there are no boundary states if the term proportional to $B$ dominates the gap, while a pair of Majorana bound states appears in the case of a $\Delta$-dominated gap. We show below that the physics of the model is much richer.

To model open boundary conditions, we make the hypothesis that the model emerges from the linearization of a spinless parabolic dispersion\cite{noise,open} with p-wave superconductivity parametrized by $\Delta$, and a resonant external field parametrized by $B$ (see SM). The condition for having a resonant field is that it has a substantial component with wavevector $2k_F$\cite{meng}. We hence get that the Fermi field $\psi(x)$ of the theory is decomposed as $\psi(x)=e^{ik_Fx}\psi_R(x)+e^{-ik_Fx}\psi_L(x)$. The open boundary conditions can be written as\cite{open}
\begin{eqnarray}
\psi_L(x)&=&-\psi_R(-x),\\
\psi_R(x+2L)&=&\psi_R(x).
\end{eqnarray}
Note that the fields $\psi_R(x)$ and $\psi_L(x)$ are not independent anymore. Moreover, the periodicity in space has doubled, leading to effective momenta $q=n\pi/L$, with $n$ integer. The Hamiltonian (1) is given by
\begin{equation}
H=\int_{-L}^L dx \left[\mathcal{H}_{0}+\mathcal{H}_{\Delta}+\mathcal{H}_B\right],
\end{equation}
with
\begin{eqnarray}
\mathcal{H}_0&=&v_F\psi^\dag_R(x)\left(-i\partial_x\right)\psi_R(x),\\
\mathcal{H}_B&=&-iB\,\mathrm{sgn}(x)\psi_R^\dag(x)\psi_R(-x),\\
\mathcal{H}_\Delta&=&i\frac{\Delta}{2}\mathrm{sgn}(x)\left[\psi_R^\dag(x)\psi_R^\dag(-x)+\psi_R(x)\psi_R(-x)\right],\end{eqnarray}
where $\mathrm{sgn}(\cdot)$ is the sign function. For the mapping from the quadratic dispersion of the common p-wave superconductor shown in the SM to the linearized model to be meaningful, band curvature at the chemical potential must give a negligible contribution to the kinetic energy. This condition holds true for large chemical potential.

Solving the Schr\"odinger equation for the problem amounts to recast the Hamiltonian in the form $H=\sum_p \epsilon_p c^\dagger_p c_p$, where $p$ is an index for a complete basis of eigenfunctions, $\epsilon_p$ is the corresponding energy, and $c_p$ the Fermi operator. To do so, we make the ansatz
\begin{equation}
c^\dag_p=\int_{-L}^L dx \left[ \chi_p(x)\psi^\dag_R(x)+\xi_p(x)\psi_R(x) \right].
\end{equation}
We then obtain the following system of differential equations
\begin{eqnarray}
\epsilon_p\chi_p(x)=-iv_F\partial_x\chi_p(x)+i\left[\Delta\xi_p(-x)-B\chi_p(-x)\right]\mathrm{sgn}(x),\nonumber\\
\epsilon_p\xi_p(x)=-iv_F\partial_x\xi_p(x)+i\left[\Delta\chi_p(-x)-B\xi_p(-x)\right]\mathrm{sgn}(x).\nonumber
\end{eqnarray}
Despite the non-local character of the equations, an analytical solution is possible. The method we employ is based on the decomposition
\begin{eqnarray}
\chi_p(x)&=&\chi_p^+(x)\mathrm{\Theta}(x)+\chi_p^-(-x)\mathrm{\Theta}(-x),\nonumber\\
\xi_p(x)&=&\xi_p^+(x)\mathrm{\Theta}(x)+\xi_p^-(-x)\mathrm{\Theta}(-x),
\end{eqnarray}
where $\mathrm{\Theta}(\cdot)$ is the Heaviside step function.\\
The additional conditions to be satisfied are $\chi_p^+(0)=\chi_p^-(0)$, $\xi_p^+(0)=\xi_p^-(0)$, $\chi_p^+(L)=\chi_p^-(L)$, $\xi_p^+(L)=\xi_p^-(L)$. Solutions are found for energies $\epsilon\geq |\Delta-B|$ and $\epsilon=0$. The energy levels in the part of the spectrum for which $\epsilon\geq |\Delta-B|$ become dense in the $L\rightarrow\infty$ limit.  The zero energy eigenfunction, henceforth labelled by a subscript $0$, represents an isolated solution. We find for the zero energy state
\begin{eqnarray}
\chi^+_0(x)&=&\chi^-_0(x)\!\!=\!\!A_0e^{(\Delta-B)x/v_F}\!+\!C_0e^{-(\Delta+B)x/v_F},\\
\xi^+_0(x)&=&\xi^-_0(x)\!\!=\!\!A_0e^{(\Delta-B)x/v_F}\!-\!C_0e^{-(\Delta+B)x/v_F}.
\end{eqnarray}
Up to a global phase, the coefficients obey
\begin{eqnarray}
A_0^2&=&C^2\frac{\Delta-B}{\Delta+B}e^{-2\Delta L/v_F}\frac{\sinh\left[(\Delta+B\right)L/v_F]}{\sinh\left[(\Delta-B\right)L/v_F]},\\
C_0^2&=&\frac{\Delta+B}{4v_F}\frac{1}{1-e^{-2(\Delta+B)L/v_F}}.
\end{eqnarray}
There are two intriguing facts about the zero energy solution. The first one is that it can be found in both of the gapped regions. This is not what is commonly expected in models for Majorana fermions, for instance, in the Kitaev model, where non-trivial boundary states only appear in the topological sector. The second observation is that for $\Delta>B$ the state is localized in the vicinity of $x=0$ and $x=L$, while for $\Delta<B$ the state is only localized close to $x=0$. The first case is the usual Majorana bound state scenario, where the fermionic zero mode is decomposed into two Majorana fermions located at the edges of the system. The second case is reminiscent of a Jackiw-Rebbi fermionic state, where the mass has a single kink. Upon varying $B$ and $\Delta$, our model implements the transition of a Jackiw-Rebbi into two Majorana bound states. This crossover is illustrated in Fig.1c).

How can this happen? To proceed with a physical interpretation, it is useful to enumerate the ingredients leading to the phenomena we have discussed. The presence of four (dependent) Fermi fields, with linear kinetic energy and zero chemical potential is needed. Furthermore, two mass terms acting in different subspaces, relations that implement a dependence between right and left movers and an 'unfolding' periodic boundary condition play essential roles. The required number of Fermi fields is provided by a helical edge proximitized by an s-wave superconductor. The boundary conditions are then implemented by two strong magnetic barriers at $x=0$ and $x=L$. The two masses are provided by the induced superconductivity and by an external magnetic field\cite{losssup,timm,invisible}. More specifically, the external magnetic field must be positive in the $-\sigma_y$ direction in spin space, the magnetic barrier at $x=0$ must be positive in the $\sigma_y$ direction in spin space, while the magnetic barrier at $x=L$ must be parallel to the external magnetic field\cite{str9}. Other directions of the magnetization of the barriers would result in twisted boundary conditions instead of Eqs.(3,4). For a schematic see Fig.1d). This analogy completely clarifies the obtained results: When the gap is of superconducting type, two Majorana fermions are present at the boundaries. On the other hand, when the gap is of magnetic type, a Jackiw-Rebbi charge is trapped close to $x=0$ since there the mass (the magnetization of the barrier/the magnetic field) changes sign. 
The mapping of the model onto a heterostructure based on the edges of a two dimensional topological insulator not only provides a valuable tool for understanding the transmutation of the Jackiw-Rebbi charge into Majorana fermions. It also provides a possible experimental realization of the model and implies that the standard techniques used to addressed the transport properties of topological heterostructures  can be employed in the case of finite magnetic barriers.

\begin{figure}
	\includegraphics[scale=1]{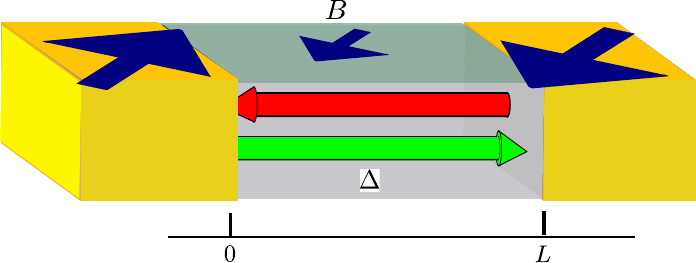}
	\caption{The quantum spin Hall analogy of the model. The arrows in the central region indicate right and left moving particles, the arrows in the side blocks the magnetization of the barriers, $B$ and $\Delta$ the applied magnetic field and the superconducting pairing.}
	\label{Fig:map}
\end{figure}
While the phase adiabatically connected to $\Delta=0$ is well understood, the Majorana phase needs to be better characterized. In particular, we now investigate the $B=0$ regime. In this case, with reference to Eq.(2), the Pauli matrices $\sigma$ become unessential, and hence the Hamiltonian density can be written as a 2x2 differential quadratic operator. One has $H=\int_0^L \Psi^\dag(x)\mathcal{H}(x)\Psi(x)dx$, with $\mathcal{H}=-iv_F\partial_x \tau_z-{i\Delta}\tau_x$. Correspondingly, the Fermi spinor acquires two components only.  As a first step, we state the inverse of Eq. (10), that reads
\begin{eqnarray}
\psi_R(x)=\sum_p \left[ \xi^*_p(x) c_p^\dag+\chi_p(x) c_p\right].
\end{eqnarray}
The explicit form of the functions $\xi_p(x)$ and $\chi_p(x)$ is given in the SM. By using Eqs. (3,4), we define the Majorana field operators $\gamma_1(x)$ and $\gamma_2(x)$ in the usual way\cite{maj1}
\begin{eqnarray}
\gamma_1(x)&=&i\left(\Psi^\dag(x)-\Psi(x)\right),\\
\gamma_2(x)&=&\left(\Psi^\dag(x)+\Psi(x)\right).
\end{eqnarray}
The zero energy contributions $\gamma^{(0)}_1$, $\gamma^{(0)}_2$ to the Majorana fields, that is the Majorana zero modes, then read
\begin{eqnarray}
\gamma^{(0)}_1&=&\frac{2\sin(k_Fx)\sqrt{\Delta/v_F}}{\sqrt{1-e^{-2\Delta L/v_F}}}e^{-\Delta(L-x)/v_F}(c^\dagger_0+c_0),\\
\gamma^{(0)}_2&=&\frac{-2i\sin(k_Fx)\sqrt{\Delta/v_F}}{\sqrt{1-e^{-2\Delta L/v_F}}}e^{-\Delta x/v_F}(c^\dagger_0-c_0).
\end{eqnarray}
We recover the expected results, namely, that one Majorana zero mode is located close to $x=0$ ($\gamma^{(0)}_2$) and one close to $x=L$ ($\gamma^{(0)}_1$). Moreover, $2k_F$ oscillations appear in accordance with the fact that we have imposed a sharp confinement potential\cite{composite}. Note that, within the model, the two Majorana modes do not hybridize.\\

Another feature of the model is that the Green functions can be calculated analytically. This allows us to show explicitly the intimate connection between the spatial extension of the Majorana zero modes given in Eqs. (18,19) and the odd-frequency component of the superconducting pairing that characterizes the topological superconductor.\\
We define the retarded Green function $G^R_{ij}(x,x',\omega)$ as\cite{flensberg}
\begin{equation}
G^R_{ij}(x,x',\omega)=\int_{-\infty}^{\infty} e^{i\omega(t+i0^+)} G^R_{ij}(x,x',t)dt,
\end{equation}
with
\begin{equation}
G^R_{ij}(x,x',t)=-i\theta(t)\langle\left\{\Psi_i(x,t),\Psi_j^\dag(x',0)\right\}\rangle,
\end{equation}
where $\Psi_i(x,t)$, ($i=1,2$) are the components of the Nambu spinor in the Heisenberg picture. The average is performed on the ground state and the braces indicate the anticommutator. The advanced Green function $G^A_{ij}(x,x',\omega)$ is given by $G^{A*}_{ij}(x,x',\omega)=G^R_{ji}(x',x,\omega)$.
Due to particle-hole symmetry of the BdG Hamiltonian, the components of the Green functions are not independent, but satisfy
\begin{equation}
G^R_{ij}(x,x',\omega)=-\sigma_x^{il} G^{R*}_{lm}(x,x',-\omega)\sigma_x^{mj}.
\end{equation}
Moreover, focussing on the the anomalous part of the Green function, that is, on the off-diagonal parts, we have $G^{A}_{21}(x',x,\omega)=-G^{R}_{21}(x,x',-\omega)$.
The function $\mathcal{F}(x,x',\omega)=G^R_{21}(x,x',\omega)+G^A_{21}(x,x',\omega)$ hence satisfies
\begin{equation}
\mathcal{F}(x,x',\omega)+\mathcal{F}(x',x,-\omega)=0.
\end{equation}
Consequently,
\begin{equation}
\mathcal{F}_+(x,x',\omega)=\frac{\mathcal{F}(x,x',\omega)+\mathcal{F}(x',x,\omega)}{2}
\end{equation}
is odd in $\omega$ and characterizes the odd-frequency pairing. Odd-frequency pairing can be expected to be related to the Majorana wave function, because a Majorana zero mode is an intrinsically odd-frequency object\cite{str15}. In our model, we find, for $\omega<\Delta$,
\begin{equation}
\mathcal{F}_+(x,x,\omega)= \frac{4{\pi}P\left(\frac{1}{\omega}\right)}{\sinh[L\sqrt{\Delta^2-\omega^2}]}\zeta(x),
\end{equation}
where $P(\cdot)$ is the Cauchy principal value and
\begin{equation}
\zeta(x)=\sin^2(k_Fx)\sinh\left[(L-2x)\sqrt{\Delta^2-\omega^2}/v_F\right].
\end{equation}
For $(L-2x)/L\simeq 1$ ($(L-2x)/L\simeq -1$), that is, close to the edges of the system, the odd-frequency pairing resembles the modulus square of the spatial extension of $\gamma^{(0)}_2$ ($\gamma^{(0)}_1$). This intriguing dependence has recently been numerically analysed in the Kitaev chain\cite{tanakak}.

In the remainder, we address the question whether the Majorana wavefunction and the odd-frequency pairing can be measured. The answer is related to the diagonal part $G^R_{11}(x,x,\omega)$ of the Green function. As an example of a measurable quantity that can be extracted from the retarded Green function, we in fact analyse the tunneling density of states $\rho(x,\omega)=-\mathrm{Im}G^R_{11}(x,x,\omega)/\pi$, that is associated with the tunneling from a metallic tip at position $x$ above the topological superconductor\cite{flensberg}. The explicit result for the Green function, for $\omega<\Delta$, reads (see also Ref.\cite{zazu2})
\begin{eqnarray}
G^R_{11}(x,x,\omega)=\frac{\omega}{\sqrt{\Delta^2-\omega^2}}\left[F(\omega,x)-F(\omega,0)\right]+\\
\frac{2}{\omega}\sqrt{\Delta^2-\omega^2}F(\omega,x)\sin^2(k_Fx)-\nonumber\\
\sin(2k_Fx)\frac{\sinh\left[(L-2x)\sqrt{\Delta^2-\omega^2}/v_F\right]}{\sinh\left(L\sqrt{\Delta^2-\omega^2}/v_F\right)},\nonumber
\end{eqnarray}
with 
\begin{equation}
F(\omega,x)=\frac{\cosh\left[\sqrt{\Delta^2-\omega^2}(L-2x)/v_F\right]}{\sinh\left(L\sqrt{\Delta^2-\omega^2}/v_F\right)}.
\end{equation}
The full Green function does not show a pronounced similarity with the Majorana wavefunction and the odd-frequency pairing. However, we find that
\begin{equation}
\rho(x,\omega)=2\Delta\delta(\omega)\sin^2(k_Fx)\frac{\cosh\left[\Delta(L-2x)/v_F\right]}{\sinh\left(\Delta L/v_F\right)}.
\end{equation}
The tunneling density of states has hence the same short wavelength components as the odd-frequency pairing, that is $\sin^2(k_Fx)$, while its envelope function is given by the derivative of the function enveloping the odd-frequency pairing.\\

In conclusion, we have proposed and analytically solved a model that is characterized by a transition between a state hosting a single Jackiw-Rebbi soliton and a state with two unpaired Majorana fermions. We have explained the results on the basis of a hybrid system involving topological edge channels, superconductivity, and magnetic gaps. We have then characterized the Majorana phase of the system on the basis of the correlation functions. We have shown that the odd-frequency component of the pairing closely follows the spatial extension of the Majorana bound states. After computing the retarded Green function, we have proposed that the tunneling density of states indeed provides information about the Majorana wave function and the odd-frequency pairing.
\begin{acknowledgements}
We thank Felix Keidel for useful discussions. This work was supported by the DFG (SPP1666, SFB1170 “ToCoTronics”),
the W\"urzburg-Dresden Cluster of Excellence ct.qmat,
EXC2147, project-id 39085490, the Elitenetzwerk
Bayern Graduate School on “Topological insulators” and the Studienstiftung des Deutschen Volkes.
\end{acknowledgements}

	\title{Supplementary material to "From fractional solitons to Majorana fermions in a model of topological superconductivity"}
	\author{N. Traverso Ziani}
	\affiliation{Dipartimento di Fisica, Universit\`a di Genova, 16146 Genova, Italy}
	\affiliation{CNR spin, 16146 Genova, Italy}
	\email{traversoziani@fisica.unige.it}
	\author{C. Fleckenstein}
	\affiliation{Institute of Theoretical Physics and Astrophysics, University of W\"urzburg, 97074 W\"urzburg, Germany}
	\author{L. Vigliotti}
	\affiliation{Dipartimento di Fisica, Universit\`a di Genova, 16146 Genova, Italy}
	\author{B. Trauzettel}
	\affiliation{Institute of Theoretical Physics and Astrophysics, University of W\"urzburg, 97074 W\"urzburg, Germany}
	\affiliation{W\"urzburg-Dresden Cluster of Excellence ct.qmat, Germany}
	\author{M. Sassetti}
	\affiliation{Dipartimento di Fisica, Universit\`a di Genova, 16146 Genova, Italy}
	\affiliation{CNR spin, 16146 Genova, Italy}
	\maketitle
	\section{Linearization of the finite p-wave superconductor}
	The starting point is a spinless p-wave superconductor, with Hamiltonian ($\hbar=1$)
	\begin{widetext}
		\begin{equation}
		H_{2}=\int_0^L dx \left[\psi^\dag(x)\left(-\frac{\partial_x^2}{2m^*}-\mu+B_2\sin(2k_F x)\right)\psi(x)+ \left(\Delta_2 \psi^\dag(x)\partial_x\psi^\dag(x)+\mathrm{h.c.}\right)\right].
		\end{equation}
	\end{widetext}
	In Eq.(1), $\psi$ is the Fermi operator, $\mu$ the chemical potential, $k_F=\sqrt{2\mu m^*}$ the Fermi momentum, $m^*$ the effective mass, $\Delta_2$ the p-wave pairing potential, and $B_2$ a competing mass. The term proportional to $B_2$ can emerge for example due to the interplay with phonons\cite{p1,p2}, or can be artificially engineered by external gates capacitively coupled to the system\cite{p3,p4}. For large enough chemical potential, we can safely perform the linearization of the theory around the Fermi points, identifying $v_F=\sqrt{2\mu/m^*}$ as Fermi velocity. With periodic boundary conditions, the diagonalization is straightforward. With open boundary conditions for the fermionic operator $\psi(0)=\psi(L)=0$, the procedure is more cumbersome but similar to Ref.\cite{open}. One has
	\begin{equation}
	\psi(x)=e^{ik_Fx}\psi_R(x)+e^{-ik_Fx}\psi_L(x),
	\end{equation}
	with $\psi_R(x)$ and $\psi_L(x)$ obeying the boundary conditions reported in the main text in Eqs.(4,5). Explicitly, 
	\begin{equation}
	\psi_R(x)=\frac{-i}{\sqrt{2L}}\sum_{n=-\infty}^{\infty}e^{in\pi x/L}c_{(n+L)k_F/\pi},
	\end{equation}
	with $c_p$ fermionic operator annihilating an electron with wavefunction $\zeta(x)=\sqrt{2/L}\sin(p\pi x/L)$. By neglecting fast oscillating terms, in the limit $Lk_F/\pi\ll 1$, and upon renormalization of the parameters ($B=B_2/2$, $\Delta=\Delta_2 k_F$), the Hamiltonian in Eq.(6) of the main text is recovered. One comment is in order: Since $\Delta\sim k_F\Delta_2$ and the linearization is only valid for large $k_F$, in view of the fact that the gap in the spinless topological superconductor diverges as the chemical potential tends to infinity, the linear model only describes the properties of the original quadratic model in the limit of large gap. This means that little or no hybridization of the Majorana modes is expected.
	\section{Eigenfunctions}
	We provide the explicit expression for the eigenfunctions in Eq.(11) of the main text that correspond to non-zero energy, in the case $B=0$. From the eigenfunctions, the Green functions are calculated analytically.\\
	We find
	\begin{eqnarray}
	\chi_q^{(+)}&=&A^{(q)}_{+}e^{i\pi q\ x/L}+B_{+}^{(q)}e^{-i\pi qx/L},\\
	\chi_q^{(-)}&=&A^{(q)}_{-}e^{i\pi qx/L}+B_{-}^{(q)}e^{-i\pi qx/L},\\
	\xi_q^{(+)}&=&C^{(q)}_{+}e^{i\pi qx/L}+D_{+}^{(q)}e^{-i\pi qx/L},\\
	\xi_q^{(-)}&=&C^{(q)}_{-}e^{i\pi qx/L}+D_{-}^{(q)}e^{-i\pi qx/L},
	\end{eqnarray}
	where $q$ is a positive integer. The corresponding excitation energies are $\epsilon_q=\sqrt{v_F^2\pi^2q^2/L^2+\Delta^2}$. The solutions for negative energy $\epsilon_q=-\sqrt{v_F^2q^2+\Delta^2}$ are the complex conjugate of the negative of the solutions with positive eigenvalue. This fact can be directly inferred from the symmetries of the Hamoltonian in Eq.(1), with $B=0$.\\
	For $q$ positive and even, we obtain the coefficients
	\begin{eqnarray}
	A_+^{(q)}&=&\frac{\epsilon_q+v_F\pi q/L}{\sqrt{8L}\epsilon_q},\\
	A_-^{(q)}&=&-\frac{\epsilon_q-v_F\pi q/L}{\epsilon_q+v_F\pi q/L}A_+,\\
	B_+^{(q)}&=&A_-^{(q)},\\
	B_-^{(q)}&=&A_+^{(q)},\\
	C_+^{(q)}&=&C_-^{(q)},\\
	C_-^{(q)}&=&-\frac{i\Delta}{\sqrt{8L}\epsilon_q},\\
	D_+^{(q)}&=&-C_+^{(q)},\\
	D_-^{(q)}&=&-C_-^{(q)}.
	\end{eqnarray}
	For $q$ odd, we have to replace $A_{\pm}^{(q)}\leftrightarrow C_{\pm}^{(q)}$, $B_{\pm}^{(q)}\leftrightarrow D_{\pm}^{(q)}$.


\begin{thebibliography}{99}
\bibitem{maj1}Y. Oreg, G. Refael, and F. von Oppen, Phys. Rev. Lett. \textbf{105}, 177002 (2010).
\bibitem{maj2}R. M. Lutchyn, J. D. Sau, and S. Das Sarma, Phys. Rev. Lett. \textbf{105}, 077001 (2010).
\bibitem{maj3}V. Mourik, K. Zuo, S. M. Frolov, S. R. Plissard, E. P. A. M. Bakkers, and L. P. Kouwenhoven, Science \textbf{336}, 1003 (2012).
\bibitem{maj4}M. T. Deng, S. Vaitiekenas, E. B. Hansen, J. Danon,
M. Leijnse, K. Flensberg, J. Nyg˚ard, P. Krogstrup, and
C. M. Marcus, Science \textbf{354}, 1557 (2016).
\bibitem{maj5}S. M. Albrecht, A. P. Higginbotham, M. Madsen, F. Kuemmeth, T. S. Jespersen, J. Nygrd, P. Krogstrup, and C. M. Marcus, Nature \textbf{531}, 206 (2016).
\bibitem{mqcanyons1}A. Y. Kitaev, Physics-Uspekhi 44, 131 (2001).
\bibitem{mqcanyons2}C. Nayak, S. H. Simon, A. Stern, M. Freedman, S. D. Sarma, Rev. Mod. Phys. \textbf{80}, 1083 (2008).
\bibitem{planar1}A. Fornieri, A. M. Whiticar, F. Setiawan, E. Portolés Marín, A. C. C. Drachmann, A. Keselman, S. Gronin, C. Thomas, T. Wang, R. Kallaher, G. C. Gardner, E. Berg, M. J. Manfra, A. Stern, C. M. Marcus, and F. Nichele, Nature \textbf{569}, 89 (2019).
\bibitem{planar2}H. Ren, F. Pientka, S. Hart, A. Pierce, M. Kosowsky, L. Lunczer, R. Schlereth, B. Scharf, E. M. Hankiewicz, L. W. Molenkamp, B. I. Halperin, and A. Yacoby, Nature \textbf{569}, 93 (2019).
\bibitem{missings}
J. Wiedenmann, E. Bocquillon, R. S. Deacon, S. Hartinger, O. Herrmann, T. M. Klapwijk, L. Maier, C. Ames, C. Br\"une, C. Gould, A. Oiwa, K. Ishibashi, S. Tarucha, H. Buhmann, and L. W. Molenkamp, Nat. Comm. \textbf{7}, 10303 (2016).
\bibitem{aw}C. Fleckenstein, N. Traverso Ziani, A. Calzona, M. Sassetti, and B. Trauzettel, arXiv:2001.03475.
\bibitem{fmm}S. Nadj-Perge, I. K. Drozdov, J. Li, H. Chen, S. Jeon, J. Seo, A. H. MacDonald, B. A. Bernevig, and A. Yazdani, Science \textbf{346}, 602 (2014).
\bibitem{tbs1}C. Fleckenstein, F. Dominguez, N. Traverso Ziani, and B. Trauzettel
Phys. Rev. B \textbf{97}, 155425 (2018).
\bibitem{tbs2}C. Moore, T.D. Stanescu, and S. Tewari, Phys. Rev. B
\textbf{97}, 165302 (2018).
\bibitem{dis1}D. Bagrets and A. Altland, Phys. Rev. Lett. \textbf{109}, 227005
(2012).
\bibitem{dis2}J. Liu, A.C. Potter, K.T. Law, and P.A. Lee, Phys. Rev.
Lett. \textbf{109}, 267002 (2012).
\bibitem{jj1}H.J. Kwon, K. Sengupta, and V.M. Yakovenko, Eur.
Phys. J. B \textbf{37}, 349 (2004).
\bibitem{jj2}J. Michelsen, V.S. Shumeiko, and G. Wendin, Phys. Rev.
B \textbf{77}, 184506 (2008).
\bibitem{jj3}A. Zazunov, S. Plugge, and R. Egger, Phys. Rev. Lett. \textbf{121}, 207701 (2018).
\bibitem{jj4}C. K. Chiu and S. Das Sarma, Phys. Rev. B \textbf{99}, 035312 (2019).
\bibitem{nonloc}E. Prada, R. Aguado, and P. San-Jose, Phys. Rev. B \textbf{96}, 085418 (2017).
\bibitem{noise}T. Jonckheere, J. Rech, A. Zazunov, R. Egger, A. Levy Yeyati, and T. Martin, Phys. Rev. Lett. \textbf{122}, 097003 (2019).
\bibitem{jr}R. Jackiw and C. Rebbi, Phys. Rev. D \textbf{13}, 3398 (1976).
\bibitem{fr1}X.-L. Qi, T. L. Hughes, and S.-C. Zhang, Nat. Phys. 4, 273
(2008).
\bibitem{fr2}J. I. V\"ayrynen and T. Ojanen, Phys. Rev. Lett. \textbf{107}, 166804 (2011).

\bibitem{fr3}J. Klinovaja, P. Stano, and D. Loss,
Phys. Rev. Lett. \textbf{109}, (2012).
\bibitem{fr4} C. Fleckenstein, N. Traverso Ziani, and B. Trauzettel, Phys.
Rev. B \textbf{94}, 241406 (2016).

\bibitem{stable}S. Kivelson and J. R. Schrieffer, Phys. Rev. B \textbf{25}, 6447 (1982).
\bibitem{ssh}A. J. Heeger, S. Kivelson, J. R. Schrieffer, and W.-P. Su,
Rev. Mod. Phys. \textbf{60}, 781 (1988).
\bibitem{pm1}S. Gangadharaiah, L. Trifunovic, and D. Loss, Phys. Rev. Lett. \textbf{108}, 136803 (2012).
\bibitem{str1}Y. Tanaka, Y. Tanuma, and A.A. Golubov, Phys. Rev. B \textbf{76}, 054522 (2007).
\bibitem{str2}Y. Tanaka, A.A. Golubov, S. Kashiwaya, and M. Ueda, Phys. Rev. Lett. \textbf{99}, 037005 (2007).
\bibitem{str3}M. Eschrig, T. L\"ofwander, T. Champel, J.C. Cuevas, J. Kopu, and G. Sch\"on, J. Low Temp. Phys. \textbf{147}, 457 (2007).
\bibitem{str4}L. Fu, and C.L. Kane, Phys. Rev. Lett. \textbf{100}, 096407 (2008).
\bibitem{str5}L. Fu, and C.L. Kane, Phys. Rev. B \textbf{79}, 161408 (2009).
\bibitem{str6}Y. Tanaka, M. Sato, and N. Nagaosa, J. Phys. Soc. Jpn. 81, 011013 (2012).
\bibitem{str7}A.M. Black-Schaffer, and A.V. Balatsky, Phys. Rev. B \textbf{86}, 144506 (2012).
\bibitem{str8}Y. Asano, and Y. Tanaka, Phys. Rev. B \textbf{87}, 104513 (2013).
\bibitem{str9}G. Dolcetto, N. Traverso Ziani, M. Biggio, F. Cavaliere, and M. Sassetti
Phys. Rev. B \textbf{87}, 235423 (2013).
\bibitem{str10}G. Dolcetto, N. Traverso Ziani, M. Biggio, F. Cavaliere, and M. Sassetti
physica status solidi (RRL)–Rapid Research Letters \textbf{7}, 1059 (2013).
\bibitem{str11}S. Hart, H. Ren, T. Wagner, P. Leubner, M. M\"uhlbauer, C. Br\"une, H. Buhmann, L.W.
Molenkamp, and A. Yacoby, Nat. Phys. \textbf{10}, 638 (2014).
\bibitem{str12}F. Crepin, P. Burset, and B. Trauzettel, Phys. Rev. B \textbf{92}, 100507(R) (2015).
\bibitem{str13}L. Arrachea and F. von Oppen
Physica E \textbf{74}, 596 (2015).
\bibitem{str14}F. Keidel, P. Burset, and B. Trauzettel
Phys. Rev. B \textbf{97}, 075408 (2018).
\bibitem{str15}F. Dominguez, O. Kashuba, E. Bocquillon, J. Wiedenmann, R. S. Deacon, T.M.
Klapwijk, G. Platero, L.W. Molenkamp, B. Trauzettel, and E.M. Hankiewicz, Phys. Rev.
B \textbf{95}, 195430 (2017).
\bibitem{str16}J. Pico-Cortes, F. Dominguez, and G. Platero, Phys. Rev. B \textbf{96}, 125438 (2017).
\bibitem{str17}O. Kashuba, B. Sothmann, P. Burset, and B. Trauzettel, Phys. Rev. B \textbf{95}, 174516 (2017).
\bibitem{str18}J. Cayao and A.M. Black-Schaffer, Phys. Rev. B \textbf{96}, 155426 (2017).
\bibitem{str19}C. Fleckenstein, N. Traverso Ziani, and B. Trauzettel, Phys. Rev. B 97, 134523 (2018).
\bibitem{soc}P. Streda and P. Seba, Phys. Rev. Lett. \textbf{90}, 256601 (2003).
\bibitem{meng}T. Meng and D.Loss, 	Phys. Rev. B \textbf{88}, 035437 (2013).
\bibitem{open}M. Fabrizio and A. O. Gogolin
Phys. Rev. B \textbf{51}, 17827 (1995).
\bibitem{losssup}D. L. Maslov, M. Stone, P. M. Goldbart, and D. Loss, Phys.
Rev. B \textbf{53}, 1548 (1996).
\bibitem{timm}C. Timm, Phys. Rev. B \textbf{86} (15), 155456 (2012).
\bibitem{invisible}C. Fleckenstein , F. Keidel, B. Trauzettel and N. Traverso
Ziani, Eur. Phys. J. Spec. Top. \textbf{227}, 1377 (2018).
\bibitem{composite}J. Klinovaja and D. Loss, Phys. Rev. B \textbf{86}, 085408 (2012).
\bibitem{flensberg}H. Bruus and K. Flesberg, \textit{Many-body Quantum Theory in Condensed Matter} (Physics Oxford University Press, Oxford, 2004).
\bibitem{tanakak}D. Takagi, S. Tamura, and Y. Tanaka, 	arXiv:1809.09324.
\bibitem{zazu2}A. Zazunov, R. Egger, and A. Levy Yeyati, Phys. Rev. B \textbf{94},
014502 (2016).
\end{thebibliography}

\begin{thebibliography}{99}
		\bibitem{p1}H. Fr\"ohlich, Proc. Roy. Soc. A \textbf{223}, 296 (1954).
		\bibitem{p2}R. E. Peierls, \textit{Quantum Theory of Solids} ( Clarendon, Oxford, 1955 ).
		\bibitem{p3}S. Gangadharaiah, L. Trifunovic, and D. Loss, Phys. Rev. Lett. \textbf{108}, 136803 (2012).
		\bibitem{p4}M. Malard, G. I. Japaridze, and H. Johannesson, Phys. Rev. B \textbf{94}, 115128 (2016).
		\bibitem{open}M. Fabrizio and A. O. Gogolin Phys. Rev. B \textbf{51}, 17827 (1995).
	\end{thebibliography}
\end{document}